\documentclass[11pt, a4paper]{article}
\usepackage{fullpage}
\usepackage{amsmath,amssymb,amsthm,mathtools,cite,comment,graphicx}
\usepackage{hyperref}

\hypersetup{hidelinks}

\theoremstyle{plain}
\newtheorem{theorem}{Theorem}[section]
\newtheorem{lemma}[theorem]{Lemma}

\newtheorem{claim}[theorem]{Claim}
\newtheorem{corollary}[theorem]{Corollary}

\theoremstyle{definition}

\newtheorem{question}[theorem]{Question}

\theoremstyle{remark}
\newtheorem{remark}[theorem]{Remark}

\newcommand{\cP}{\mathcal{P}}
\newcommand{\cS}{\mathcal{S}}
\newcommand{\bF}{\mathbf{F}}
\newcommand{\bM}{\mathbf{M}}
\DeclareMathOperator{\pf}{pf}

\title{Exact Matching in Matrix Multiplication Time}
\author{
Ryotaro Sato\thanks{Preferred Networks, Inc., Tokyo, Japan. \texttt{sryotaro@preferred.jp}} \and
Yutaro Yamaguchi\thanks{Osaka University, Osaka, Japan. \texttt{yutaro.yamaguchi@ist.osaka-u.ac.jp}}
}
\date{\empty}

\begin{document}
\maketitle
\thispagestyle{empty}
\begin{abstract}
Let $A_0,A_1\in\bF^{n\times n}$ be square matrices over a finite field $\bF$ and consider the matrix pencil $A_0+yA_1$ with indeterminate $y$.
We observe that, once $A_0+\lambda A_1$ is nonsingular for some $\lambda\in\bF$, the polynomial $\det(A_0+yA_1)$ can be reconstructed by computing one determinant, one inverse matrix, and the characteristic polynomial of a single matrix.
Consequently, this determinant polynomial can be computed in $\mathrm{O}(n^\omega)$ field operations, avoiding the polylogarithmic overhead of a general polynomial-matrix determinant algorithm in this special setting.

Applying this observation to random evaluation of the Tutte matrix of a graph, we obtain a matrix-multiplication-time randomized algorithm for the so-called exact matching problem.
Specifically, one can decide, simultaneously for all $k$, whether a given $0/1$-weighted graph has a perfect matching of weight exactly $k$ in $\mathrm{O}(n^\omega)$ field operations, where $n$ denotes the number of vertices in the graph.
We also discuss the analogous extension to the exact linear matroid parity problem and its consequences for a perfect packing of Mader's $\mathcal{S}$-paths of minimum total length and for a shortest cycle through three specified vertices.

\end{abstract}

\newpage
\pagenumbering{roman}
\tableofcontents
\newpage
\pagenumbering{arabic}
\setcounter{page}{1}

\section{Introduction}
Throughout the paper, let $\bF$ be a field and suppose that multiplication of two matrices in $\bF^{n\times n}$ can be done in $\mathrm{O}(n^\omega)$ field operations for some $\omega>2$.
Initiated by Strassen~\cite{Strassen1969}, there have been numerous improvements, and the current bound is $\omega<2.371339$~\cite{ADWXXZ2025}.

A \emph{matching} in a graph is an edge set in which no two edges share an end vertex, and it is \emph{perfect} if it covers all vertices.
Finding a perfect (or maximum) matching is a fundamental combinatorial optimization problem.
In addition to combinatorial deterministic algorithms~\cite{Edmonds1965a,MV1980,Vazirani2024,GK2004}, a long line of work has developed randomized algorithms~\cite{MVV1987,RV1989,MS2004,Harvey2009} based on the algebraic characterization of the existence of a perfect matching using the \emph{Tutte matrix} (Theorem~\ref{thm:Tutte}).
The following matrix-multiplication-time (MM-time, for short) result is due to Mucha and Sankowski~\cite{MS2004}; Harvey~\cite{Harvey2009} subsequently gave a particularly direct algebraic formulation for nonbipartite matching.

\begin{theorem}[Mucha and Sankowski~\cite{MS2004}; see also Harvey~\cite{Harvey2009}]\label{thm:perfect}
Given a graph on $n$ vertices, one can test whether it has a perfect matching, with high probability, in $\mathrm{O}(n^\omega)$ field operations.
Furthermore, if the returned answer is yes, one can construct a perfect matching deterministically in $\mathrm{O}(n^\omega)$ field operations.
\end{theorem}

A graph with edge weight $0$ or $1$ on each edge is called a \emph{$0/1$-weighted graph}.
The \emph{exact matching problem}~\cite{PY1982} asks, given such a graph and an integer $k$, whether there is a perfect matching of weight exactly $k$.
Randomized polynomial-time algorithms initiated by Mulmuley, Vazirani, and Vazirani~\cite{MVV1987} are known, while it remains a major open problem whether this problem admits a deterministic polynomial-time algorithm.

Related algebraic work includes pseudopolynomial-time algorithms for matroid problems~\cite{CGM1992, CLL2014}, and MM-time algorithms for weighted bipartite matching~\cite{Sankowski2009}.
The standard algebraic characterization expresses the answer for every $k$ as the nonvanishing of a coefficient of the Pfaffian of the weighted Tutte matrix (Theorem~\ref{thm:MVV}).
After random substitution for the edge indeterminates, the target matrix is a polynomial matrix with a single indeterminate, for which one may compute the Pfaffian with the aid of a general polynomial-matrix determinant algorithm of Storjohann~\cite{Storjohann2003}.
This gives the following baseline, which appears to be implicit rather than explicitly stated in the earlier exact-matching literature.

\begin{theorem}[Immediate from Theorem~\ref{thm:MVV} and Corollary~\ref{cor:compute_pfaffian_general}]\label{thm:exact}
Given a $0/1$-weighted graph on $n$ vertices, one can test simultaneously for every $k=0,1,\dots,\frac{n}{2}$ whether there exists a perfect matching of weight exactly $k$, with high probability, in $\mathrm{O}(n^\omega\cdot\mathrm{poly}(\log n))$ field operations.
\end{theorem}

\subsection{Our Result}
In this paper, we show the following theorem.

\begin{theorem}\label{thm:main1}
Given a $0/1$-weighted graph on $n$ vertices, one can test simultaneously for every $k=0,1,\dots,\frac{n}{2}$ whether there exists a perfect matching of weight exactly $k$, with high probability, in $\mathrm{O}(n^\omega)$ field operations.
For each $k$ that is determined as feasible, one can construct a perfect matching of weight exactly $k$ in $\mathrm{O}(n^{\omega}\cdot\min\{n, 1 + \min\{k, \frac{n}{2} - k\} \log n\})$ field operations.
\end{theorem}

The main contribution of Theorem~\ref{thm:main1} is to remove the polylogarithmic factor from Theorem~\ref{thm:exact} for the simultaneous decision problem.
We also include several standard self-reduction ideas for constructing a solution, which are also randomized.
Improvements in the computational time and/or derandomization are left as open problems.

Here we sketch the idea of our main result.
The source of the extra polylogarithmic factor in Theorem~\ref{thm:exact} is that Storjohann's algorithm handles a general polynomial matrix.
The matrices arising here (after random substitution for the edge indeterminates in the Tutte matrices) are of the much more special form
\[A(y)=A_0+yA_1,\]
where $A_0$ and $A_1$ are matrices over $\mathbf{F}$ and $y$ is an indeterminate; such a polynomial matrix is called a \emph{matrix pencil}.
For $\lambda \in \mathbf{F}$, if $A(\lambda)=A_0+\lambda A_1$ is nonsingular, then one can rewrite its determinant by setting $s\coloneqq y-\lambda$ as follows:
\[\det A(y) =\det A(\lambda)\cdot s^n \cdot\det\!\left(s^{-1}I+A(\lambda)^{-1}A_1\right).\]
Thus, after reversing the coefficients, $\det A(y)$ is obtained from the characteristic polynomial of the matrix $-A(\lambda)^{-1}A_1$ over $\bF$.
We state and prove this identity as Lemma~\ref{lem:matrix-pencil}.
It is valid over every field and isolates the precise reason that the polylogarithmic overhead can be removed.

The same argument is not specific to matching.
We also formulate it for the \emph{exact linear matroid parity problem} and propose an algorithm for the analogous simultaneous decision problem that is comparable to the randomized algorithm \cite[Theorem~1.2]{CLL2014} for the unweighted problem. 

\begin{theorem}[Informal; the formal version appears as Theorem~\ref{thm:exact_LMP} in Section~\ref{sec:matrix_LMP}]\label{thm:exact_LMP_informal}
Given a $0/1$-weighted linear matroid parity instance represented by an $r \times 2m$ matrix over $\bF$, one can test simultaneously for every $k=0,1,\dots,\frac{r}{2}$ whether there exists a parity base of weight exactly $k$, with high probability, in $\mathrm{O}(mr^{\omega - 1})$ field operations.
For each $k$ that is determined as feasible, one can construct a parity base of weight exactly $k$ in $\mathrm{O}(r^\omega \cdot \min\{m, 1 + \min\{r, k, \frac{r}{2} - k\} \log m\})$ field operations.
\end{theorem}

We then combine it with the reduction of the shortest disjoint $\mathcal{S}$-paths problem to weighted linear matroid parity~\cite{Yamaguchi2016}.
When each edge has unit length, this gives an MM-time algorithm for computing the optimal value, which is comparable again to the randomized algorithm \cite[Theorem~1.3]{CLL2014} for the unweighted problem. 
Furthermore, as a direct application, the minimum length of a cycle through three specified vertices in an undirected graph can be computed within the same bound.

\begin{theorem}[Informal; the formal version appears as Theorem~\ref{thm:S-packing_randomized} in Section~\ref{sec:Mader}]\label{thm:S-packing_randomized_informal}
Given an instance of disjoint $\cS$-paths problem on an $n$-vertex graph, one can compute the minimum total length of a collection of vertex-disjoint $\cS$-paths covering all the terminals in $\cS$, with high probability, in $\mathrm{O}(n^\omega)$ field operations.
Under the condition that the computed value is correct, one can construct a collection of $\cS$-paths attaining the minimum in $\mathrm{O}(n^{\omega+1})$ field operations.
\end{theorem}

\begin{corollary}\label{cor:cycle}
Given an undirected graph on $n$ vertices with three specified vertices, one can compute the minimum length of a cycle through all the specified vertices, with high probability, in $\mathrm{O}(n^\omega)$ field operations.
Under the condition that the computed value is correct, one can construct a cycle attaining the minimum in $\mathrm{O}(n^{\omega+1})$ field operations.
\end{corollary}

\begin{remark}\label{rem:LasVegas}
All decision algorithms in this paper are one-sided Monte Carlo algorithms based on random evaluation of polynomial matrices: an existing solution may be missed by evaluating a nonzero polynomial to zero, whereas any nonzero coefficient certifies the existence of the corresponding solution.
The failure probability can be made $n^{-c}$ for any constant $c$ by working over a field of polynomially larger size.
All construction algorithms can be made Las Vegas algorithms with the stated computational time bounds in expectation; one can deterministically verify the correctness of the obtained solution, and repeat the procedure until a correct solution is obtained without any failure (cf.~\cite[Section~3.5]{Harvey2009}).
\end{remark}

\subsection{Organization}
The remainder of this paper is organized as follows.
Section~\ref{sec:preliminaries} collects the required algebraic facts, states the main lemma explicitly, and reviews the relation between perfect matchings and the Tutte matrices.
Section~\ref{sec:main} proves Theorem~\ref{thm:main1} and explains standard self-reduction constructions.
Section~\ref{sec:LMP} gives the exact linear matroid parity extension and its applications.

\section{Preliminaries}\label{sec:preliminaries}
\subsection{Polynomials and Matrices}
For an indeterminate $x$, let $\bF[x]$ denote the ring of polynomials in $x$ with coefficients in $\bF$.
For a set $X=\{x_1,x_2,\dots,x_m\}$ of indeterminates, we similarly write $\bF[X]=\bF[x_1,x_2,\dots,x_m]$.
Hereafter, we use $\equiv$ for equality as polynomials.
For $f\in\bF[X]$ and $x\in X$, the notation $[x^k]f$ denotes the coefficient of $x^k$, viewed as a polynomial in the remaining indeterminates; thus
\[f \equiv \sum_k ([x^k]f) \cdot x^k.\]

Let $A=(a_{i,j})$ be an $n\times n$ matrix over $\bF$ or over a polynomial ring over $\bF$.
Its determinant is
\[\det A \equiv \sum_{\sigma} \left\{\ \mathrm{sgn}(\sigma)\prod_{i=1}^{n} a_{i, \sigma(i)} \ \biggm| \ \sigma \colon \text{permutation of $[n] \coloneqq \{1, 2, \dots, n\}$} \ \right\}.\]
We use the following standard algorithmic facts.

\begin{lemma}\label{lem:determinant}
Given $A\in\bF^{n\times n}$, one can compute $\det A$ deterministically in $\mathrm{O}(n^\omega)$ field operations.
If $A$ is nonsingular, one can compute $A^{-1}$ within the same bound.
\end{lemma}

\begin{theorem}[Storjohann~\cite{Storjohann2003}]\label{thm:Storjohann}
There exists a Las Vegas algorithm such that, given $A\in\bF[x]^{n\times n}$ whose entries have degree at most $d$, it computes $\det A\in\bF[x]$ in $\mathrm{O}(n^\omega d \cdot \mathrm{poly}(\log n + \log d))$ field operations in expectation.
\end{theorem}

For $A\in\bF^{n\times n}$, its \emph{characteristic polynomial} is $\chi_A(t)\coloneqq\det(tI-A) \in \bF[t]$.
By definition, the degree of $\chi_A(t)$ is at most $n$.
There are several efficient algorithms for computing $\chi_A(t)$~\cite{Keller-Gehrig1985, RI2011, NP2021, PS2007}, and we use the following result.

\begin{theorem}[Neiger and Pernet~\cite{NP2021}]\label{thm:NP}
Given $A\in\bF^{n\times n}$, one can compute $\chi_A(t)$ deterministically in $\mathrm{O}(n^\omega)$ field operations.
\end{theorem}

The following lemma is the central algebraic ingredient of this paper.

\begin{lemma}\label{lem:matrix-pencil}
Let $A_0,A_1\in\bF^{n\times n}$ and $\lambda\in\bF$.
Suppose that $A_\lambda\coloneqq A_0+\lambda A_1$ is nonsingular.
Set $C\coloneqq-A_\lambda^{-1}A_1$ and $s\coloneqq y-\lambda$.
Then,
\begin{align}\label{eq:matrix-pencil}
\det(A_0+yA_1) \equiv \det(A_\lambda)\cdot s^n \cdot \chi_C(s^{-1}).
\end{align}
Consequently, $\det(A_0+yA_1)\in\bF[y]$ can be computed deterministically in $\mathrm{O}(n^\omega)$ field operations.
\end{lemma}

\begin{proof}
We have
\[A_0+yA_1\equiv A_\lambda+sA_1\equiv A_\lambda(I+sA_\lambda^{-1}A_1),\]
and therefore
\begin{align*}
\det(A_0+yA_1)
 &\equiv\det(A_\lambda)\cdot\det(I+sA_\lambda^{-1}A_1)\\
 &\equiv\det(A_\lambda)\cdot s^n \cdot\det(s^{-1}I+A_\lambda^{-1}A_1)\\
 &\equiv\det(A_\lambda)\cdot s^n \cdot\chi_C(s^{-1}),
\end{align*}
which proves~\eqref{eq:matrix-pencil}.

The expression $\chi_C(s^{-1})$ is interpreted in the Laurent-polynomial ring $\bF[s,s^{-1}]$; after multiplying $s^n$, it is the polynomial in $s$ displayed as follows:
\[s^n \cdot \chi_C(s^{-1})\equiv\sum_{k=0}^n c_k s^{n-k},\]
where $c_k \coloneqq [t^k]\chi_C(t)$.
Thus, the right-hand side is obtained by reversing the coefficient order of the characteristic polynomial and then translating the indeterminate from $s$ to $y-\lambda$.
The determinant and inverse of $A_\lambda$, the product $A_\lambda^{-1}A_1$, and the characteristic polynomial of $C$ can be computed deterministically in $\mathrm{O}(n^\omega)$ operations by Lemma~\ref{lem:determinant} and Theorem~\ref{thm:NP}.
The coefficient reversal and polynomial translation require only $\mathrm{O}(n^2)$ operations, which completes the proof.
\end{proof}

While Theorem~\ref{thm:Storjohann} applies to arbitrary polynomial matrices and is therefore more general, Lemma~\ref{lem:matrix-pencil} focuses on matrix pencils and reduces the main task to one ordinary characteristic-polynomial computation.
This is exactly where the improvement from $\mathrm{O}(n^\omega \cdot \mathrm{poly}(\log n))$ to $\mathrm{O}(n^\omega)$ arises.

We next recall Pfaffians.
An $n\times n$ matrix $A$ is \emph{alternating} if $A^\top=-A$ (\emph{skew-symmetric}) and every diagonal entry is zero; the diagonal condition is relevant when the characteristic of $\mathbf{F}$ is $2$.
Suppose that $n$ is even and $A$ is alternating.
Its Pfaffian is defined as
\begin{align*}
\pf A \coloneqq \sum_{\sigma} \left\{\ \mathrm{sgn}(\sigma)\prod_{i=1}^{n/2} a_{\sigma(2i - 1), \sigma(2i)} \ \Biggm| \ \begin{array}{ll}\sigma \colon \text{permutation of $[n]$},\\[0mm] \sigma(1) < \sigma(3) < \cdots < \sigma(n - 1),\\[0mm] \sigma(2i - 1) < \sigma(2i) \ (i = 1, 2, \dots, \frac{n}{2}) \end{array} \ \right\}.
\end{align*}
Equivalently, this is a signed sum over the perfect matchings of the complete graph on $[n]$.
The definition remains valid in the characteristic $2$ case, where the signs disappear.

The following fact is well known, which is useful to compute $\pf A$ from $\det A$.

\begin{lemma}\label{lem:pf-square}
For any even-dimensional alternating matrix $A$ over a commutative ring,
\[(\pf A)^2\equiv\det A.\]
\end{lemma}

To recover the Pfaffian from its square, it suffices to compute square roots of field elements that are promised to be squares.

For a finite field $\bF$, let $R_{\bF}$ denote the number of field operations used by a square-root routine on a promised square.
Suppose that the characteristic of $\mathbf{F}$ is $2$, i.e., $\bF=\mathbb{F}_{2^q}$ for some integer $q$.
Then, the Frobenius map $a\mapsto a^2$ is an automorphism (as $(a + b)^2 = a^2 + b^2$), and the square root of any $b \in \mathbb{F}_{2^q}$ is uniquely given by $b^{2^{q-1}}$ as its inverse.
Thus, we have $R_{\bF}=\mathrm{O}(q)=\mathrm{O}(\log|\bF|)$ by iterated squaring.

\begin{lemma}\label{lem:pfaffian}
Let $A\in\bF[x]^{n\times n}$ be an even-dimensional alternating matrix over a finite field $\bF$.
Given $g\equiv\det A \in \bF[x]$, one can recover a polynomial $h\in \bF[x]$ with $h\equiv\pf A$ or $h\equiv-\pf A$ in $\mathrm{O}(D^2+D R_{\bF})$ field operations, where $D=\deg g$.
In particular, when the characteristic of $\mathbf{F}$ is $2$, we have $h\equiv\pf A$ and the bound is $\mathrm{O}(D\log|\bF|)$ deterministically.
\end{lemma}

\begin{proof}
By Lemma~\ref{lem:pf-square}, $g\equiv h^2$ for $h\equiv \pf A$.
If $g\equiv 0$, it suffices to output $h \equiv 0$.
Assume first that the characteristic of $\mathbf{F}$ is not $2$.
Write $h=\sum_jh_jx^j$, and let $r$ be the smallest index with $h_r\ne0$.
Then the least nonzero term of $g$ has degree $2r$ and coefficient $h_r^2$.
Use the scalar square-root routine to choose one of the two possibilities for $h_r$.
For $k=r+1,r+2,\dots$, comparison of the coefficient of $x^{r+k}$ gives
\[[x^{r+k}]g = 2h_rh_k + \sum_{j = 1}^{k - r - 1}h_{r+j}h_{k-j}.\]
All terms in the sum in the right-hand side have already been determined, and $2h_r\ne0$, so this equation determines $h_k$ uniquely.
Choosing the other square root for $h_r$ just changes the sign of every coefficient.
There are $\mathrm{O}(D)$ coefficients and each comparison uses $\mathrm{O}(D)$ operations.
Thus, the required operations are $\mathrm{O}(D^2 + R_\bF)$ in total.

Now suppose that the characteristic of $\mathbf{F}$ is $2$.
Then,
\[\left(\sum_jh_jx^j\right)^2=\sum_jh_j^2x^{2j}.\]
Hence, every odd-degree coefficient of $g$ is zero and $h_j$ is the unique square root of $[x^{2j}]g$.
Applying the inverse Frobenius map coefficient-wise, i.e., $h_j = ([x^{2j}]g)^{|\bF|/2}$, proves the stated deterministic bound.
\end{proof}

Combining Theorem~\ref{thm:Storjohann} with Lemma~\ref{lem:pfaffian} gives the following corollary.
It is only used to justify the baseline Theorem~\ref{thm:exact}.
In the proof of Theorem~\ref{thm:main1}, Lemma~\ref{lem:matrix-pencil} replaces Theorem~\ref{thm:Storjohann}.

\begin{corollary}\label{cor:compute_pfaffian_general}
There exists a Las Vegas algorithm such that, given alternating $A\in\bF[x]^{n\times n}$ whose entries have degree at most $d$, where $\bF$ is a finite field of size $n^{\mathrm{O}(1)}$, 
it computes $\pf A \in \bF[x]$ up to a global sign in $\mathrm{O}(n^\omega d \cdot \mathrm{poly}(\log n+\log d) +(nd)^2+ndR_{\bF})$ field operations in expectation.
\end{corollary}

\begin{proof}
Theorem~\ref{thm:Storjohann} computes the determinant within the first term, which has degree $D \le nd$.
Lemma~\ref{lem:pfaffian} then recovers its polynomial square root within the remaining terms.
\end{proof}

\subsection{Matching and Tutte Matrix}
Let $G = (V, E)$ be a simple graph.
We introduce an indeterminate $x_e$ for each edge $e \in E$, and let $X_E \coloneqq \{\, x_e \mid e \in E \,\}$ denote the set of those indeterminates.
The \emph{Tutte matrix} $T(G)$ of $G$ is an alternating matrix in $\bF[X_E]^{V \times V}$ defined as follows, where we fix a total order $<$ on $V$: 
\begin{align*}
    T(G)_{u, v} \coloneqq \begin{cases}
        x_e & \text{if $e = \{u, v\} \in E$ and $u < v$,}\\
        -x_e & \text{if $e = \{u, v\} \in E$ and $u > v$,}\\
        0 & \text{otherwise.}
    \end{cases}
\end{align*}

\begin{theorem}[Tutte~\cite{Tutte1947}]\label{thm:Tutte}
    A graph $G$ has a perfect matching if and only if $\det T(G) \not\equiv 0$.
\end{theorem}

We turn to the exact matching problem.
Let $G_0 = (V, E_0)$ and $G_1 = (V, E_1)$ be two simple graphs on the same vertex set $V$, and let $G = G_0 + G_1 = (V, E = E_0 \dot\cup E_1)$ be the disjoint union of $G_0$ and $G_1$.
That is, $G$ has no selfloop but may have at most two parallel edges between each pair of distinct vertices, one from $G_0$ and the other from $G_1$.
For each subset $F \subseteq E$, we define its \emph{weight} as $|F \cap E_1|$.
The \emph{exact matching problem}~\cite{PY1982} asks the existence of a perfect matching whose weight is exactly $k$.

We introduce an extra indeterminate $y$ and extend the Tutte matrix as follows.
The \emph{Tutte matrix} of $(G_0, G_1)$ is an alternating matrix in $\bF[X_E, y]^{V \times V}$ defined as $T(G_0, G_1) \coloneqq T(G_0) + yT(G_1)$.

\begin{theorem}[cf.~\cite{MVV1987, CGM1992}]\label{thm:MVV}
    For each $k = 0, 1, \dots, \frac{n}{2}$, a graph $G = G_0 + G_1$ on $n$ vertices has a perfect matching of weight exactly $k$ if and only if $[y^k]\pf T(G_0, G_1) \not\equiv 0$.
\end{theorem}

\subsection{An $\mathrm{O}(n^\omega)$-Time Randomized Algorithm for Perfect Matching}\label{sec:PM}
We review a very simple, $\mathrm{O}(n^\omega)$-time randomized algorithm to test the existence of a perfect matching in a graph on $n$ vertices, which proves the first part of Theorem~\ref{thm:perfect}.
The algorithm is based on Theorem~\ref{thm:Tutte} and the following theorem, so-called the \emph{Schwartz--Zippel Lemma}.

\begin{theorem}[cf.~{\cite[Theorem~7.2]{MR1995}}]\label{thm:SZ}
    Let $f \in \bF[x_1, x_2, \dots, x_m]$ be a polynomial with indeterminates $x_1, x_2, \dots, x_m$ of total degree $d$ such that $f \not\equiv 0$.
    Let $S \subseteq \bF$ be a finite subset, and choose $r_1, r_2, \dots, r_m \in S$ independently and uniformly at random.
    Then, 
    \begin{align*}
        \mathrm{Pr}\left(f(r_1, r_2, \dots, r_m) = 0\right) \le \frac{d}{|S|}.
    \end{align*}
\end{theorem}

Let $G = (V, E)$ be a simple graph with $|V| = n$.
By definition, the total degree of $\det T(G)$ is at most $n$.
Choose $q = \mathrm{O}(\log n)$ so that $|\bF|=2^q \gg n^2$, and work over the binary extension field $\bF=\mathbb{F}_{2^q}$.
For each edge $e \in E$, choose $r_e \in \bF$ independently and uniformly at random.
Let $\tilde{T}(G) \in \bF^{V \times V}$ denote the matrix obtained from $T(G)$ by substituting $r_e$ for $x_e$ $(e \in E)$.

By Theorems~\ref{thm:Tutte} and \ref{thm:SZ}, if $G$ has a perfect matching, then
\[\mathrm{Pr}\left(\det \tilde{T}(G) = 0\right) \le \frac{n}{|\bF|} \ll \frac{1}{n},\]
and otherwise $\det \tilde{T}(G) = 0$.
Thus, by Lemma~\ref{lem:determinant}, we can test in $\mathrm{O}(n^\omega)$ time whether $G$ has a perfect matching or not, with high probability.

After we confirm that $\det\tilde{T}(G) \neq 0$, a perfect matching can be constructed deterministically by a simple self reduction as follows.
Note that the elementary edge-by-edge reduction described here does not give the $\mathrm{O}(n^\omega)$ bound stated in the second part of Theorem~\ref{thm:perfect}.
The MM-time implementation uses divide and conquer together with fast low-rank updates of inverse matrices; see \cite{MS2004, Harvey2009}.

First, $\det \tilde{T}(G) \neq 0$ is equivalent to $\pf \tilde{T}(G) \neq 0$ by Lemma~\ref{lem:pf-square}.
Pick any edge $e \in E$, and let $G_e$ and $G^e$ be the graphs obtained from $G$ by removing $e$ (meaning that $e$ is forced to be unused) and by removing all the edges intersecting $e$ except for $e$ itself (meaning that $e$ is forced to be used), respectively. 
Also, let $\tilde{T}(G_e)$ and $\tilde{T}(G^e)$ be the corresponding matrices obtained from $\tilde{T}(G)$ by replacing $r_e$ with $0$ and by replacing $r_{e'}$ with $0$ for all the edges $e'$ intersecting $e$ except for $e$ itself, respectively.
Then, by the definition of Pfaffian (focusing on $r_e$), we have
\[0 \neq \pf \tilde{T}(G) = \pf \tilde{T}(G_e) + \pf \tilde{T}(G^e),\]
which means that at least one of $\det \tilde{T}(G_e)$ and $\det \tilde{T}(G^e)$ is nonzero.
Thus, one can reduce the instance $G$ to $G_e$ or $G^e$ just by removing edges and updating the matrix accordingly.
Proceeding through all edges, we obtain a perfect matching.

\section{An $\mathrm{O}(n^\omega)$-Time Randomized Algorithm for Exact Matching}\label{sec:main}
In this section, we prove Theorem~\ref{thm:main1}.
Let $G=G_0+G_1$ be a $0/1$-weighted graph on $n$ vertices, with edge sets $E_0$ and $E_1$.
We first run the perfect-matching test in Section~\ref{sec:PM}.
If the test reports that $G$ has no perfect matching, we report that every weight is infeasible; this is correct unless the preliminary test incurs its one-sided error.
Otherwise, the nonzero determinant found by that test certifies that $G$ has a perfect matching.
We assume this from now on, and reuse the same value $r_e$ for random evaluation of each $x_e$.

\subsection{Simultaneous Decision for All Weights}\label{sec:decision}
Fix a constant $c>0$.
Choose $q = \mathrm{O}(\log n)$ so that $|\bF|=2^q\ge n^{c+2}$, and work over the binary extension field $\bF=\mathbb{F}_{2^q}$.
For each edge $e\in E \coloneqq E_0\mathbin{\dot\cup}E_1$, choose $r_e\in\bF$ independently and uniformly at random.
Let $A(y) \coloneqq A_0+yA_1\in\bF[y]^{V\times V}$ be the matrix obtained from $T(G_0,G_1)$ by substituting $r_e$ for $x_e$ $(e \in E)$, where $A_i$ is the evaluated Tutte matrix of $G_i$.
The matrix $A(y)$ is clearly alternating, as is $T(G_0, G_1)$.

For each $k$, define the polynomial in the edge indeterminates
\[f_k(X_E)\coloneqq[y^k]\pf T(G_0,G_1).\]
By Theorem~\ref{thm:MVV}, $f_k\not\equiv0$ exactly characterizes when $G$ has a perfect matching of weight $k$.
Since the total degree of $f_k$ is at most $\frac{n}{2}$, by Theorem~\ref{thm:SZ},
\[\Pr\bigl([y^k]\pf A(y)=0 \mid f_k\not\equiv0\bigr) \le \frac{n}{2|\bF|} \le \frac{1}{n^{c+1}}.\]
The union bound over the at most $\frac{n}{2}+1$ feasible values of $k$ shows that all nonzero coefficients survive simultaneously with probability at least $1-\mathrm{O}(n^{-c})$.

Now we utilize Lemma~\ref{lem:matrix-pencil} to compute $\pf A(y)$.
At $y=1$, the matrix $A(1)=A_0+A_1$ coincides with the evaluated Tutte matrix $\tilde{T}(G)$ of the unweighted graph $G$ in Section~\ref{sec:PM}.
By the assumption that the preliminary test has passed, $A(1)$ is nonsingular.
Thus, we can apply Lemma~\ref{lem:matrix-pencil} with $\lambda=1$.
Let $C\coloneqq -(A_0+A_1)^{-1}A_1$ and $s \coloneqq y-1$.
Then,
\begin{align*}\label{eq:exact-pencil}
\det A(y) \equiv \det(A_0+A_1) \cdot s^n \cdot \chi_C(s^{-1}),
\end{align*}
which can be computed deterministically in $\mathrm{O}(n^\omega)$ field operations: compute the determinant and inverse of $A_0+A_1$, form $C$, compute $\chi_C(t)$, reverse its coefficients, and translate $s$ back to $y-1$.
By Lemma~\ref{lem:pfaffian}, we then recover $\pf A(y)$ in $\mathrm{O}(n \log n)$ further field operations (recall that we employ the field $\bF$ of characteristic $2$ with $|\bF| = 2^q$ and $q = \mathrm{O}(\log n)$).
Reading all coefficients $[y^k]\pf A(y)$ proves the decision part of Theorem~\ref{thm:main1}.

\subsection{Self Reduction for Exact Matching}\label{sec:self-reduction}
We next explain the construction bound.
Fix a value $k$ that the decision procedure has declared feasible.
We use the preceding $\mathrm{O}(n^\omega)$-time procedure for Pfaffian computation as an oracle for subgraphs, where random substitution for the edge indeterminates $x_e$ $(e \in E)$ is fixed at the beginning but $\lambda = 1$ for Lemma~\ref{lem:matrix-pencil} may be replaced by a random value at some point.

Let $H=H_0+H_1$ be the current graph, initially $G_0+G_1$, and $A^{H}(y) \equiv A^H_0 + yA^H_1 \in \bF[y]^{V \times V}$ be the corresponding matrix obtained by the fixed substitution to $T(H_0, H_1)$.
We maintain the invariant that $[y^k]\pf A^H(y) \neq 0$, which implies that $H$ has a perfect matching of weight $k$ by Theorem~\ref{thm:MVV}.

We show two types of algorithms:
one is for the general case and requires $\mathrm{O}(n)$ iterations, and the other is for the case where $k$ or $\frac{n}{2} - k$ is small and requires $\mathrm{O}(\min\{k, \frac{n}{2} - k\} \log n)$ iterations.

\subsubsection{General $\mathrm{O}(n)$-iteration algorithm}\label{sec:self-reduction_general}
For a vertex $v$ and $i \in \{0, 1\}$, let $\delta_i^H(v)$ be the set of weight-$i$ edges of $H$ (i.e., in $H_i$) incident to $v$.
By definition (focusing on the vertex $v$), we have
\[[y^k]\pf A^{H - \delta_1^H(v)}(y) +  [y^k]\pf A^{H - \delta_0^H(v)}(y) = [y^k]\pf A^{H}(y) \neq 0.\]
This implies at least one of the two terms in the left-hand side is nonzero, which enables us to remove one of $\delta_0^H(v)$ and $\delta_1^H(v)$ from $H$ while preserving the invariant.
Specifically, we compute $\pf A^{H - \delta_i^H(v)}(y)$ $(i = 0, 1)$ in $\mathrm{O}(n^\omega)$ time using the oracle.

Here, $A^{H - \delta_i^H(v)}(1)$ may be singular even if $\pf A^{H - \delta_i^H(v)}(y) \not\equiv 0$.
Then, in order to apply Lemma~\ref{lem:matrix-pencil}, we need to pick a random value $\lambda \in \bF$ making $A^{H - \delta_i^H(v)}(\lambda)$ nonsingular with high probability (by Theorem~\ref{thm:SZ}).
Since the invariant guarantees that $[y^k]\pf A^{H - \delta_i^H(v)}(y) \neq 0$ for at least one of $i = 0, 1$, by alternately trying to compute them, we obtain the results $\det A^{H - \delta_i^H(v)}(\lambda) \neq 0$ and $[y^k]\pf A^{H - \delta_i^H(v)}(y) \neq 0$ for some $i$ within expected $\mathrm{O}(1)$ trials.
This makes the self reduction a Las Vegas algorithm even with a single execution.

If $[y^k]\pf A^{H - \delta_0^H(v)}(y) \neq 0$, then we replace $H$ by removing $\delta_0^H(v)$ from $H_0$, which forces $v$ to be matched with a weight-$1$ edge.
Otherwise, we have $[y^k]\pf A^{H - \delta_1^H(v)}(y) \neq 0$ and replace $H$ by removing $\delta_1^H(v)$ from $H_1$, which forces $v$ to be matched with a weight-$0$ edge.
Proceed through all vertices.

At the end, each vertex is matched by an edge of the same weight in any perfect matching.
Consequently, every perfect matching has the same weight, which is exactly $k$ by the maintained invariant.
Thus, it suffices to construct an arbitrary perfect matching in the final graph by the second part of Theorem~\ref{thm:perfect}.
There are $n$ oracle calls (applications of Lemma~\ref{lem:matrix-pencil} each with expected $\mathrm{O}(1)$ trials of random selection of $\lambda$), each taking $\mathrm{O}(n^\omega)$ field operations, so the total is $\mathrm{O}(n^{\omega+1})$.
This self-reduction construction itself is a Las Vegas algorithm as seen above.

Performing this separately for all feasible $k$, one can find perfect matchings of all possible weights in expected $\mathrm{O}(n^{\omega+2})$ field operations in total.
This leaves the following question.

\begin{question}[cf.~\cite{Yamaguchi2025}]\label{qu:reconstruction}
Is there a faster and/or deterministic construction algorithm?
On the computational time, natural target bounds, in increasing order of strength, are the following:
\begin{enumerate}
\setlength{\itemsep}{0.5mm}
\item $\mathrm{O}(n^{\omega+1})$ field operations for all $k$ in total;
\item $\mathrm{O}(n^\omega)$ field operations for one specified feasible $k$;
\item $\mathrm{O}(n^\omega)$ field operations for all $k$ in total.
\end{enumerate}
\end{question}

\subsubsection{$\mathrm{O}(k \log n)$-iteration algorithm for small $k$}\label{sec:self-reduction_small}
When $k$ is small, it suffices to fix $k$ edges of weight $1$ to be included in the target perfect matching.
Similarly, when $\frac{n}{2} - k$ is small, it suffices to fix $\frac{n}{2} - k$ edges of weight $0$ to be included.
Here we focus on the former case with $k > 0$ (when $k = 0$, just find a perfect matching in $G_0$).

In this case, we keep $H_0 = G_0$, and perform a binary search to find a weight-$1$ edge that must be included to form a perfect matching of weight exactly $k$.
We remark that it is not clear whether a single execution of this self reduction can be made a Las Vegas algorithm with the same expected bound like the previous one; nevertheless, one can make it Las Vegas by repeating the same procedure until a correct solution is obtained (cf.~Remark~\ref{rem:LasVegas}).

Fix an ordering $e_1, e_2, \dots, e_\ell$ of the edges in $H_1$, and let $A_{1}^{H, j}$ $(j = 0, 1, \dots, \ell)$ be the prefix sums of their contributions to the evaluated Tutte matrix $\tilde{T}(H_1)$, i.e., $A_{1}^{H, j} \coloneqq A_1^{H - \{e_{j+1}, e_{j+2}, \dots, e_\ell\}}$.
In particular, $A_{1}^{H,0} = O$ and $A_{1}^{H,\ell} = A_1^H$.
Let $A^{H,j}(y) \coloneqq A_0^H + yA_{1}^{H,j}$.
Then, we have
\[[y^k]\pf A^{H,0}(y) = [y^k]\pf A_0^H = 0 \neq [y^k]\pf A^H(y) = [y^k]\pf A^{H,\ell}(y),\]
which implies that there exists an index $j^*$ such that $[y^k]\pf A^{H, j^*-1}(y) = 0 \neq [y^k]\pf A^{H,j^*}(y)$.
Such $j^*$ can be found by a standard binary search starting from the interval $[0, \ell]$ with the invariant that the current interval $[L, R]$ satisfies $[y^k]\pf A^{H, L}(y) = 0 \neq [y^k]\pf A^{H,R}(y)$.
This requires $\mathrm{O}(\log n)$ iterations, and in each step we only need one oracle call in $\mathrm{O}(n^\omega)$ time.

Here, in order to apply Lemma~\ref{lem:matrix-pencil}, we need to pick a random value $\lambda \in \bF$ making $A^{H, j}(\lambda)$ nonsingular with high probability when $[y^k]\pf A^{H, j}(y) \neq 0$ (by Theorem~\ref{thm:SZ}), again.
Unlike the previous self reduction, it is nontrivial whether this can be performed absolutely correctly within the same expected computational time bound under the assumption that $[y^k]\pf A^{H, j}(y) = 0 \neq [y^k]\pf A^{H,j'}(y)$ $(j < j^* \le j')$.
This makes it unclear whether a single execution of this binary-search self reduction can be made a Las Vegas algorithm with the same expected bound.

Once such $j^*$ is found, we can conclude that the edge $e_{j^*}$ can be forced to be included in our target perfect matching of weight $k$ (as $e_{j^*}$ has a nonzero contribution to $[y^k]\pf A^H(y)$) and the edges $e_i$ $(i > j^*)$ can be removed.
Thus, we remove all the edges $e_i$ $(i > j^*)$ and all the edges intersecting $e_{j^*}$ except for $e_{j^*}$ itself, and update the ordering of the remaining edges in $H_1$ by moving $e_{j^*}$ to the top so that it will not be removed.

Repeating the same procedure $k$ times, $H$ has $k$ edges of weight $1$ that are forced to be included in any perfect matching, and the maintained invariant $[y^k]\pf A^H(y) \neq 0$ implies the existence of a perfect matching of weight $k$.
Thus, it suffices to construct an arbitrary perfect matching in the graph obtained by removing the weight-$1$ edges other than those $k$ edges, which can be done by the second part of Theorem~\ref{thm:perfect}, again.
The total required operations are bounded by $\mathrm{O}((k\log n) \cdot n^\omega)$.

\section{Extension to Linear Matroid Parity}\label{sec:LMP}
\subsection{Linear Matroid Parity Problem}
Let $Z\in\bF^{r\times U}$ be a matrix whose column set $U$ is partitioned into two-element subsets, called \emph{lines}; let $L$ denote the line set.
After elementary row operations and deletion of zero rows, we may assume that $Z$ has row-full rank $r$.
A subset $W\subseteq U$ is a \emph{parity set} if it is a union of lines.
For such a set, we write $L(W)\coloneqq\{\,\ell\in L \mid \ell\subseteq W\,\}$, and sometimes identify $W$ and $L(W)$.
The column dependencies of $Z$ define a linear matroid on the ground set $U$, denoted by $\bM(Z)$ (for the basics on matroids, see, e.g., \cite{Schrijver2003}).
A base of $\bM(Z)$ that is a parity set is called a \emph{parity base}.
In particular, a parity base can exist only when $r$ is even.

The \emph{linear matroid parity problem} asks for a maximum-cardinality independent parity set in a given pair $(Z,L)$.
It simultaneously generalizes nonbipartite matching and linear matroid intersection.
Starting with Lov\'{a}sz~\cite{Lovasz1981}, several polynomial-time algorithms have been developed, including the deterministic augmenting-path algorithm of Gabow and Stallmann~\cite{GS1986} and the randomized algebraic algorithm of Cheung, Lau, and Leung~\cite{CLL2014}.
The weighted problem asks for a minimum-weight parity base when each line has a weight.
Iwata and Kobayashi~\cite{IK2022} gave a deterministic strongly polynomial-time algorithm, while algebraic pseudopolynomial-time algorithms go back to Camerini, Galbiati, and Maffioli~\cite{CGM1992}.

Linear matroid parity has many applications through (sometimes highly nontrivial) reductions, including Mader's disjoint $\cS$-paths~\cite{Lovasz1980,Schrijver2003}, feedback vertex sets in subcubic graphs~\cite{UKG1988}, maximum-genus embeddings~\cite{FGM1988}, and parity-constrained orientations~\cite{FJS2001}.
In what follows, we give a general discussion on exact linear matroid parity as a generalization of exact matching, and then focus on Mader's disjoint $\cS$-paths problem to make explicit how our method enters the reduction.

\subsection{Matrix Formulation of Linear Matroid Parity}\label{sec:matrix_LMP}
Let $(Z,L)$ be an LMP instance with $Z\in\bF^{r\times U}$ and $|L|=m$.
For a line $\ell\in L$, let $z_{\ell,1},z_{\ell,2}\in\bF^r$ be its two columns.
For $a,b\in\bF^r$, define
\[a\wedge b\coloneqq ab^\top-ba^\top\in\bF^{r\times r}.\]
Introduce an indeterminate $x_\ell$ for each line $\ell \in L$ and define the alternating matrix
\[Y(Z,L)\coloneqq\sum_{\ell\in L}x_\ell(z_{\ell,1}\wedge z_{\ell,2}).\]

\begin{theorem}[Lov\'{a}sz~\cite{Lovasz1979}]\label{thm:lovasz-lmp}
An instance $(Z,L)$ has a parity base if and only if $\det Y(Z,L)\not\equiv0$.
\end{theorem}

This formulation naturally includes the Tutte matrix as a special case.
For each edge $e=\{u,v\}$ of a graph, take a line consisting of the two unit vectors $\mathbf{1}_u$ and $\mathbf{1}_v$ in $\bF^V$; then $Y(Z,L)$ is the Tutte matrix after a consistent choice of signs.

There are also more compact mixed-skew-symmetric and delta-matroid formulations and related path-packing structures; see, e.g., Geelen and Iwata~\cite{GeelenIwata2005}, Wahlstr\"om~\cite{Wahlstrom2024}, and Koana and Wahlstr\"om~\cite{KoanaWahlstrom2025}.
For our argument, Lov\'{a}sz' matrix formulation above is sufficient and makes the pencil structure transparent.

Now partition $L=L_0\mathbin{\dot\cup}L_1$, assigning weight $0$ to the lines in $L_0$ and weight $1$ to those in $L_1$.
For a parity set $W$, its weight is $|L(W)\cap L_1|$.
Introduce an extra indeterminate $y$ and set
\[Y(Z,L_0,L_1) \coloneqq Y(Z,L_0)+yY(Z,L_1).\]

\begin{theorem}[Camerini, Galbiati, and Maffioli~\cite{CGM1992}]\label{thm:CGM}
Let $(Z, L_0, L_1)$ be a 0/1-weighted LMP instance.
We assume that the two columns belonging to each line are arranged consecutively in $Z$. 
Then, 
\[\pf Y(Z,L_0,L_1) \equiv \sum_{B \in \binom{L}{r/2}} \det Z[B] \cdot y^{|B\cap L_1|} \prod_{\ell\in B}x_\ell,\]
where $Z[B]$ is the $r\times r$ submatrix of $Z$ obtained by arranging the two columns of every line in $B$ in the same order as in $Z$. 
Consequently, for each $k=0,1,\dots,\frac{r}{2}$, the instance $(Z, L_0, L_1)$ has a parity base of weight exactly $k$ if and only if $[y^k]\pf Y(Z,L_0,L_1)\not\equiv0$.
\end{theorem}

Now we are ready to state the formal version of Theorem~\ref{thm:exact_LMP_informal}, and prove two parts (simultaneous decision and construction) separately in the following subsections (similar to Section~\ref{sec:main}).

\begin{theorem}\label{thm:exact_LMP}
Let $(Z,L_0,L_1)$ be a $0/1$-weighted LMP instance over a finite field $\mathbf{F}$, where $Z\in\bF^{r\times U}$ is of row-full rank and $m=|L|$.
Assume $|\bF| \ge (r + m)^{c + 2}$ for some constant $c > 0$ by taking a finite extension field if necessary.
Assume also that promised scalar square roots can be computed in $\mathrm{O}(r^{\omega - 1})$ field operations; this condition holds deterministically when the characteristic of $\bF$ is $2$ and $|\bF| = 2^{\mathrm{O}(\log r)}$.
Then, one can decide simultaneously for every $k = 0, 1, \dots, \frac{r}{2}$ whether there is a parity base of weight exactly $k$, with high probability, in $\mathrm{O}(r^\omega+mr^{\omega - 1}) = \mathrm{O}(mr^{\omega - 1})$ field operations; if each column of $Z$ has at most $s$ nonzero entries, the second term can be replaced by $\mathrm{O}(ms^2)$.
For each $k$ that is determined as feasible, one can construct a parity base of weight exactly $k$ in $\mathrm{O}(r^\omega \cdot \min\{m, 1 + \min\{r, k, \frac{r}{2} - k\} \log m\})$ field operations.
\end{theorem}

\subsubsection{Simultaneous decision for all weights}
If $r$ is odd, no parity base exists, so assume that $r$ is even.
Choose $\alpha_\ell\in\bF$ independently and uniformly at random and substitute $x_\ell=\alpha_\ell$.
This gives a matrix pencil
\[A(y)\equiv A_0+yA_1, \quad\text{where}~A_i = \sum_{\ell\in L_i}\alpha_\ell(z_{\ell,1}\wedge z_{\ell,2}).\]
If the instance has no parity base, then $A(1)$ is singular for every substitution and every answer is negative.
If it has a parity base, Theorems~\ref{thm:lovasz-lmp} and \ref{thm:SZ} imply that $A(1)=A_0+A_1$ is nonsingular with high probability.
Thus, if the sampled $A(1)$ is singular, the algorithm reports that all weights are infeasible.

Suppose that $A(1)$ is nonsingular.
Then,  in $\mathrm{O}(r^\omega)$ operations, Lemma~\ref{lem:matrix-pencil} computes $\det A(y)$ and Lemma~\ref{lem:pfaffian} recovers $\pf A(y)$ (recall the assumption on the promised scalar square root computation).
For each feasible $k$, the coefficient $[y^k]\pf Y(Z, L_0, L_1)$ in Theorem~\ref{thm:CGM}, viewed as a polynomial in $x_\ell$, has total degree $\frac{r}{2}$.
The union bound over the at most $\frac{r}{2}+1$ values of $k$, together with the event that $A(1)$ is nonsingular, proves simultaneous correctness with high probability as in Section~\ref{sec:decision}.

The second term in the computational time bound comes from computation of $A_i$ $(i = 0, 1)$.
Computing $z_{\ell,1} \wedge z_{\ell,2}$ naively requires $\mathrm{O}(r^2)$ operations, while it suffices to compute for only $\mathrm{O}(s^2)$ nonzero positions when each vector has at most $s$ nonzero entries; this gives the $\mathrm{O}(ms^2)$ bound.
We can obtain the $\mathrm{O}(mr^{\omega - 1})$ bound by partitioning the $m$ lines into $\frac{m}{r}$ subsets $L^j$ of $r$ lines and computing the following for each $L^j$ using fast matrix multiplication in $\mathrm{O}(r^\omega)$ time:
\[\sum_{\ell \in L^j}\alpha_\ell (z_{\ell,1} \wedge z_{\ell,2}) = PQ^\top - QP^\top,\]
where $P, Q \in \bF^{r \times r}$ are the matrices obtained by arranging $\alpha_\ell z_{\ell, 1}, z_{\ell, 2}$ $(\ell \in L^j)$, respectively.
This completes the proof for the first part of Theorem~\ref{thm:exact_LMP}.

\subsubsection{Self reduction for exact linear matroid parity}\label{sec:construction_LMP}
Now, we consider the construction part.
The following elementary line-by-line self reduction works, which incurs an additional factor of $m$ to the first term, i.e., gives the $\mathrm{O}(mr^\omega)$ bound (cf.~\cite[Section~7]{CLL2014}).
For each line $\ell \in L$ in an arbitrary order, compute the Pfaffian after removing $\ell$ by Lemmas~\ref{lem:matrix-pencil} and \ref{lem:pfaffian}, and if the coefficient of $y^k$ remains nonzero, remove $\ell$ and update the matrix.
The union bound implies that a single scan of the lines determines a parity base of weight $k$ with high probability.
This can be simply made a Las Vegas algorithm with the same expected bound (expected $\mathrm{O}(1)$ scans are sufficient).

We can also obtain the $\mathrm{O}(r^{\omega + 1} \log m)$ bound as well as the $\mathrm{O}(1 + \min\{k, \frac{r}{2} - k\} \cdot r^{\omega} \log m)$ bound based on the binary search idea as follows. This is similar to Section~\ref{sec:self-reduction_small}, and it is unclear whether a single execution can be made Las Vegas.

First, fix an ordering $\ell_1, \ell_2, \dots, \ell_m$ of the lines in $L$, and let $A_{i}^j$ $(i = 0, 1; j = 0, 1, \dots, m)$ be the prefix sums of their contributions to the evaluated matrix, i.e.,
\[A_i^j \coloneqq \sum_{\ell \in L_i^j} \alpha_{\ell}(z_{\ell, 1} \wedge z_{\ell, 2}), \quad \text{where}~L_i^j \coloneqq L_i \cap \{\, \ell_{j'} \mid j' \le j \,\}.\]
In particular, $A_i^0 = O$ and $A_i^{m} = A_i$.
Let $A^{j}(y) \coloneqq A_0^j + yA_{1}^{j}$.
Then, we have
\[[y^k]\pf A^{0}(y) = [y^k]\pf O = 0 \neq [y^k]\pf A(y) = [y^k]\pf A^{m}(y),\]
which implies that there exists an index $j^*$ such that $[y^k]\pf A^{j^*-1}(y) = 0 \neq [y^k]\pf A^{j^*}(y)$.
Such $j^*$ can be found by a standard binary search starting from the interval $[0, m]$ with the invariant that the current interval $[L, R]$ satisfies $[y^k]\pf A^{L}(y) = 0 \neq [y^k]\pf A^{R}(y)$.
This requires $\mathrm{O}(\log m)$ iterations, and in each step we only need one oracle call in $\mathrm{O}(r^\omega)$ time.

Once such $j^*$ is found, we can conclude that the line $\ell_{j^*}$ can be forced to be included in our target parity base of weight $k$ (as $\ell_{j^*}$ has a nonzero contribution to $[y^k]\pf A(y)$) and the lines $\ell_j$ $(j > j^*)$ can be removed.
In addition, after removing $\ell_j$ $(j > j^*)$, all the parity base of weight $k$ in the remaining instance contain $\ell_{j^*}$ with high probability; this is because $[y^k]\pf A^{j^*-1}(y) = 0$ implies $[y^k]\pf Y(Z, L_0^{j^*-1}, L_1^{j^*-1}) \equiv 0$ with high probability by Theorem~\ref{thm:SZ}, which concludes that the instance $(Z, L_0^{j^*-1}, L_1^{j^*-1})$ has no parity base of weight $k$ by Theorem~\ref{thm:CGM}.
Thus, we remove $\ell_j$ $(j \ge j^*)$ with treating $\ell_{j^*}$ as included (by adding the contribution $y^i \cdot \alpha_{\ell_{j^*}}(z_{\ell_{j^*},1} \wedge z_{\ell_{j^*},2})$ to all $A^j(y)$ $(j < j^*)$, where $\ell_{j^*} \in L_i$).

Repeating the same procedure $\frac{r}{2}$ times, we obtain a parity base of weight exactly $k$.
When $k$ or $\frac{r}{2} - k$ is small, we can reduce the number of iterations to $\min\{k, \frac{r}{2} - k\}$ times in the same way as Section~\ref{sec:self-reduction_small}.
After recognizing $k$ weight-$1$ lines or $\frac{r}{2} - k$ weight-$0$ lines to be fixed, by contracting them and deleting all the remaining lines of the same weight ($1$ or $0$, respectively), we obtain a smaller $0/1$-weighted LMP instance in which all the parity bases have the same weight.
By computing a parity base in it in $\mathrm{O}(r^\omega)$ time by \cite{CLL2014} and adding back the fixed lines, we can obtain a parity base of weight exactly $k$ in the original instance.

The prefix sums $A_i^j$ $(i = 0, 1; j = 0, 1, \dots, m)$ can be computed in advance, and addition of the contributions of fixed lines can be dealt with separately.
While naive precomputation of all $A_i^j$ may require $\Theta(mr^2)$ time and space, we can reduce them to $\mathrm{O}(r^\omega)$ per query (oracle call) in addition to $\mathrm{O}(mr^{\omega - 1})$ at the beginning as follows.
We precompute $A_i^j$ only for $j = 0, r, 2r, \dots, \left\lfloor \frac{m}{r} \right\rfloor r$ in advance using fast matrix multiplication as shown at the end of the previous section.
Then, for each query, we can obtain $A_i^j$ for any $j$ by adding the contributions of less than $r$ lines to $A_i^{j'}$ with $j' = \left\lfloor \frac{j}{r} \right\rfloor r$, which can be done by performing fast matrix multiplication $\mathrm{O}(1)$ times.
This completes the proof of Theorem~\ref{thm:exact_LMP}.

\begin{remark}
Any general improvement of the self reduction part is left as an open problem as Question~\ref{qu:reconstruction}.
In the following application to Mader's $\cS$-paths problem, we improve the overhead by exploiting the fact that the matrix derives from the graph. 
\end{remark}

\subsection{Shortest Disjoint $\cS$-Paths}\label{sec:Mader}
Let $G=(V,E)$ be a simple undirected graph, let $T\subseteq V$ be a terminal set, and let $\cS$ be a partition of $T$.
A \emph{$T$-path} is a simple path between two distinct terminals whose internal vertices avoid $T$.
It is an \emph{$\cS$-path} if its two ends belong to distinct classes of $\cS$.
An \emph{$\cS$-packing} is a collection of vertex-disjoint $\cS$-paths, and it is \emph{perfect} if it covers all the terminals in $T$.
The problem of finding a maximum $\cS$-packing was first raised by Gallai~\cite{Gallai1961}, and Mader established its min-max theorem~\cite{Mader1978}.
The first polynomial-time algorithm was obtained through matroid matching~\cite{Lovasz1980,Lovasz1981}; a more direct reduction to linear matroid parity was shown by Schrijver~\cite{Schrijver2003}.

The weighted version asks for a perfect $\cS$-packing of minimum total length.
Related weighted work in edge-disjoint and terminal-pair settings includes~\cite{Karzanov1993,Karzanov1997,HP2014}.
Yamaguchi~\cite{Yamaguchi2016} reduced this problem to weighted linear matroid parity.
We now give an essential part of the reduction needed in the unit-weight case, where the length is evaluated by the number of edges.
The following lemma and its proof sketch are based on \cite[Section~2]{Yamaguchi2016}, where we translate $A$ in \cite{Yamaguchi2016} as $T$ here and set $k = \frac{|T|}{2}$.

\begin{lemma}\label{lem:reduction}
Let $(G=(V,E), T, \cS)$ be an instance of disjoint $\cS$-paths problem with $|V| = n$ and $|E| = m$.
Assume that $|T|$ is even.
Then, one can construct a $0/1$-weighted LMP instance $(Z,L_0,L_1)$ with the following properties deterministically in $\mathrm{O}(n + m)$ time:
\begin{enumerate}
\setlength{\itemsep}{0.5mm}
\item The representation matrix $Z$ is over any finite field $\bF$ with $|\bF| \ge |\cS|+2$; in particular, one may use a binary extension field.
It has $r=\mathrm{O}(n)$ rows, $m$ lines each corresponding to an edge in $E$, and $\mathrm{O}(n)$ extra lines independent of $E$, and every column of $Z$ has at most four nonzero entries.
Note that the last condition enables $\mathrm{O}(n + m)$-time implicit construction of $Z$, which is an $\mathrm{O}(n) \times \mathrm{O}(n + m)$ matrix having $\mathrm{O}(n + m)$ nonzero entries (while the original description of \cite{Yamaguchi2016} naively requires $\mathrm{O}(n(n + m))$-time explicit construction).
\item Every perfect $\cS$-packing $\mathcal{P}$ gives a parity base $B$ with $|B \cap L_1| = |E(\mathcal{P})|$.
\item Every parity base $B$ gives, deterministically in $\mathrm{O}(n)$ time, a perfect $\cS$-packing $\mathcal{P}_B$ with $|E(\mathcal{P}_B)|\le |B \cap L_1|$.
Consequently, the minimum weight of a parity base equals the minimum total length of a perfect $\cS$-packing.
\end{enumerate}
\end{lemma}

\begin{proof}[Proof Sketch]
We describe construction of the auxiliary instance $(G' = (V', E'), T', \cS')$ and correspondence between $\cS'$-packings and parity bases in the reduced LMP instance, which are necessary for the self-reduction construction of a perfect $\cS$-packing in the proof of Theorem~\ref{thm:S-packing_randomized}, while we skip the precise construction of the LMP instance, for which we refer to \cite[Section~2.1]{Yamaguchi2016}.

Let $b_1, b_2$ be two extra terminals, each forming a new singleton terminal class; that is,
\begin{align*}
    V' &\coloneqq V \cup \{b_1, b_2\},\\
    T' &\coloneqq T \cup \{b_1, b_2\},\\
    \cS' &\coloneqq \cS \cup \{\{b_1\}, \{b_2\}\}.
\end{align*}
We add an edge $e'$ between $b_1$ and $b_2$, which forms a unique $\cS'$-path with end vertex $b_2$.
We also add an edge $e_v$ between $b_1$ and each $v \in V$.
Thus,
\[E' \coloneqq E \cup \{e'\} \cup \{\, e_v \mid v \in V \,\}.\]
The following claim bridges the perfect $\cS$-packings and the parity bases in some LMP instance, which immediately implies the first property in Lemma~\ref{lem:reduction}.

\begin{claim}[cf.~{\cite[Lemma~4]{Yamaguchi2016}}]\label{cl:correspondence}
    There exists an LMP instance $(Z, L)$ with $Z \in \bF^{2|V'| \times 2|E'|}$ such that
    \begin{itemize}
    \setlength{\itemsep}{0.5mm}
        \item every column of $Z$ has at most four nonzero entries;
        \item each line corresponds to an edge in $E'$, and vice versa;
        \item a line set $B \subseteq L$ forms a parity base if and only if the edge set corresponding to $B$ forms a spanning forest of $G'$ in which every connected component contains exactly one $T'$-path and it is an $\cS'$-path.
    \end{itemize}
\end{claim}

Finally, we assign weight $1$ to the lines corresponding to the original edges in $E$, and weight $0$ to the lines corresponding to the additional edges in $E' \setminus E$.
Then, the third property in Claim~\ref{cl:correspondence} implies the second and third properties in Lemma~\ref{lem:reduction} as follows.

Let $\cP$ be a perfect $\cS$-packing in $G$.
Then, by adding $e'$ and $e_v$ for each vertex $v \in V$ not incident to any edge in $E(\cP)$, we expand $E(\cP)$ to a spanning forest of $G'$ satisfying the condition in the third property in Claim~\ref{cl:correspondence}.
Since all the added edges are in $E' \setminus E$, the weight of the corresponding parity base is clearly $|E(\cP)|$, which implies the second property in Lemma~\ref{lem:reduction}.

Conversely, let $B$ be a parity base.
Then, the third property in Claim~\ref{cl:correspondence} implies that $B$ corresponds to a spanning forest of $G'$ that includes a perfect $\cS$-packing in $G$ (recall that $b_2 \in T'$ can only be connected to $b_1 \in T'$ by its unique incident edge $e'$).
The size of a spanning forest of $G'$ is $\mathrm{O}(n)$, from which it is easy to extract the $\cS$-paths in linear time; this implies the third property in Lemma~\ref{lem:reduction}.
\end{proof}

\begin{remark}
The reduction preserves only the minimum value, not the entire set of attainable values.
Accordingly, it does not yield a polynomial-time algorithm for computing a perfect $\cS$-packing with an arbitrarily specified number of edges; that problem already contains the Hamiltonian path problem as a special case, which is NP-hard.
\end{remark}

The randomized MM-time consequence, the formal version of Theorem~\ref{thm:S-packing_randomized_informal}, now follows directly from Theorem~\ref{thm:exact_LMP}; we include an improvement on the construction in the proof.

\begin{theorem}\label{thm:S-packing_randomized}
Given an instance $(G=(V,E), T, \cS)$ of disjoint $\cS$-paths problem with $|V| = n$, one can decide whether a perfect $\cS$-packing exists, and if so, compute the minimum total length of a perfect $\cS$-packing, with high probability, in $\mathrm{O}(n^\omega)$ field operations.
Under the condition that the computed value $\mu$ is correct, one can construct a perfect $\cS$-packing of the minimum total length in $\mathrm{O}(n^{\omega} \cdot \min\{n, 1 + \min\{\mu - \frac{|T|}{2}, n - (\mu + \frac{|T|}{2})\}\log n\})$ field operations.
\end{theorem}

\begin{proof}
Construct the LMP instance of Lemma~\ref{lem:reduction}.
By the first condition, its evaluated matrices $A_0, A_1$ used in Theorem~\ref{thm:exact_LMP} can be constructed in $\mathrm{O}(n^2)$ operations, and all feasible weights of parity bases can be determined, with high probability, by computing the Pfaffian of $A(y) \equiv A_0 + yA_1$ in $\mathrm{O}(n^\omega)$ operations.
If $\pf A(y) \equiv 0$, then with high probability, there is no parity base and Lemma~\ref{lem:reduction} certifies that no perfect $\cS$-packing exists.
Otherwise, let $\mu$ be the minimum value for which $[y^\mu] \pf A(y) \neq 0$, which is the minimum total length of a perfect $\cS$-packing with high probability.

Now we discuss the construction of a perfect $\cS$-packing of total length $\mu$ under the assumption that $\mu$ is exactly the minimum total length of a perfect $\cS$-packing.
We perform the self reduction for the vertices in $V \setminus T$ (similar to that in Section~\ref{sec:self-reduction}).
We work on the auxiliary instance $(G', T', \cS')$ described in the proof sketch of Lemma~\ref{lem:reduction}.
Let $H$ be the current graph, initially $G'$, and $A^{H}(y) = A^H_0 + yA^H_1$ be the corresponding matrix obtained from $A(y)$ by replacing the values substituted for the line indeterminates corresponding to the edges in $E' \setminus E(H)$ with $0$.
We maintain the invariant that $[y^\mu]\pf A^H(y) \neq 0$, which implies that $H$ has a spanning forest of $G'$ including a perfect $\cS$-packing of total length $\mu$ in $G$ by Claim~\ref{cl:correspondence}.

We first explain a vertex-to-vertex reduction leading to the $\mathrm{O}(n^{\omega + 1})$ bound; this can be made a Las Vegas algorithm with a single execution like Section~\ref{sec:self-reduction_general}.
For a vertex $v \in V \setminus T$, let $\delta^H(v)$ be the set of edges in $E$ incident to $v$ that remain in $H$.
Recall that $e_v \in E' \setminus E$ is the extra edge between $v$ and $b_1$.
Since $\mu$ is the correct minimum value, any spanning forest $F$ of $G'$ contributing to $[y^\mu]\pf A^{H}(y)$ satisfies either $F \cap \delta^H(v) = \emptyset$ or $F \cap \{e_v\} = \emptyset$ (otherwise, it contains a perfect $\cS$-packing of total length less than $\mu$).
Thus, we have 
\[[y^\mu]\pf A^{H - \delta^H(v)}(y) + [y^\mu]\pf A^{H - e_v}(y) = [y^\mu]\pf A^{H}(y) \neq 0.\]
This implies that at least one of the two terms in the left-hand side is nonzero, which enables us to remove one of $\delta^H(v)$ and $e_v$ from $H$ while preserving the invariant.
Specifically, we compute $\pf A^{H - \delta^H(v)}(y)$ and $\pf A^{H - e_v}(y)$ using the oracle.
If $[y^\mu]\pf A^{H - \delta^H(v)}(y) \neq 0$, then we replace $H$ by removing $\delta^H(v)$; this forces $v$ to be disjoint from the target perfect $\cS$-packing (and then connected to $b_1$ by $e_v$ in a spanning forest corresponding to a parity base), where at least one perfect $\cS$-packing of total length $\mu$ satisfies this.
Otherwise, we have $[y^\mu]\pf A^{H - e_v}(y) \neq 0$ and replace $H$ by removing $e_v$; this forces $v$ to be included in the target perfect $\cS$-packing, where at least one perfect $\cS$-packing of total length $\mu$ satisfies this, again.
Proceed through all vertices.

At the end, each vertex $v \in V \setminus T$ with $\delta^H(v) \neq \emptyset$ must intersect all the perfect $\cS$-packings of total length $\mu$.
Consequently, every perfect $\cS$-packing in the final graph is of the same total length, which is exactly $\mu$ by the minimality of $\mu$.
Thus, it suffices to construct an arbitrary perfect $\cS$-packing in the final graph in $\mathrm{O}(n^\omega)$ time by \cite[Section~5.1.3]{CLL2014} (this construction part conditioned by the nonzeroness of the Pfaffian is deterministic as with Theorem~\ref{thm:perfect}).
There are $n$ oracle calls (Pfaffian computations), each taking $\mathrm{O}(n^\omega)$ field operations, so the total is $\mathrm{O}(n^{\omega+1})$.

The $\mathrm{O}(n^\omega \cdot (1 + \min\{\mu - \frac{|T|}{2}, n - (\mu + \frac{|T|}{2})\} \log n))$ bound can be derived by incorporating the binary search (as with Sections~\ref{sec:self-reduction_small} and \ref{sec:construction_LMP}) into the above argument for fixing which vertices in $V \setminus T$ are used or unused.
Note that any perfect $\cS$-packing of total length $\mu$ intersects exactly $\mu - \frac{|T|}{2}$ vertices in $V \setminus T$ and then it is disjoint from $n - |T| - (\mu - \frac{|T|}{2}) = n - (\mu + \frac{|T|}{2})$.
\end{proof}

Lemma~\ref{lem:reduction} can also be combined with the deterministic weighted LMP algorithm of Iwata and Kobayashi~\cite{IK2022}.
This yields the following previously known consequence.

\begin{theorem}[Iwata and Kobayashi~\cite{IK2022} and Yamaguchi~\cite{Yamaguchi2016}]\label{thm:S-packing_deterministic}
Given an instance $(G=(V,E), T, \cS)$ of disjoint $\cS$-paths problem with $|V| = n$ and $|E| = m$, one can decide whether a perfect $\cS$-packing exists, and if so, find a perfect $\cS$-packing of the minimum total length deterministically in $\mathrm{O}(nm^3)$ field operations.
\end{theorem}

\subsection{Shortest Cycle Through Three Specified Vertices}
Finally, we prove Corollary~\ref{cor:cycle}, where the problem is to determine the length of a shortest cycle through three specified vertices.
For a more general setting, Björklund, Husfeldt, and Taslaman~\cite{BHT2012} gave an FPT algorithm parameterized by the number of specified vertices.
Other algebraic formulations are known as well~\cite{Wahlstrom2013,EKW2024,EibenKoanaWahlstrom2025}.

\begin{theorem}[Björklund, Husfeldt, and Taslaman~\cite{BHT2012}]
Given a simple undirected graph $G=(V,E)$ and a terminal set $T\subseteq V$, one can compute the length of a shortest cycle through all vertices in $T$, with high probability, in $\mathrm{O}(2^{|T|}n^3)$ field operations.
A cycle of the computed length can be constructed deterministically in $\mathrm{O}(2^{|T|}n^4\log n)$ field operations.\footnote{This can be reduced to $\mathrm{O}(2^{|T|}n^4)$ by improving the self reduction part: instead of the binary search written in the paper~\cite{BHT2012}, one can decide the next transition by solving the current instance once from one of the end vertices to all the neighbors of the other end vertex.}
\end{theorem}

For $|T|=3$, the reduction to a perfect $\cS$-packing is particularly simple.
Let $T=\{t_1,t_2,t_3\}$.
Replace each $t_i$ by two copies $t_i^+,t_i^-$, each adjacent to every former neighbor of $t_i$.
Let
\[T'\coloneqq\{\, t_i^+,t_i^- \mid i=1,2,3 \,\}, \quad \cS'\coloneqq\{\,\{t_i^+,t_i^-\} \mid i=1,2,3\,\}.\]

A perfect $\cS'$-packing consists of three paths.
Regardless of pairing of the end vertices, one can obtain a simple cycle through $t_1,t_2,t_3$ in the original graph by merging back each pair $t_i^+,t_i^-$ into $t_i$ (since the three paths are internally vertex-disjoint).
Conversely, such a cycle decomposes into three internally disjoint arcs; assigning the two incidences at each terminal to its two copies gives a perfect $\cS'$-packing.
Both transformations preserve the total length.

Thus, combining this length-preserving correspondence with Theorem~\ref{thm:S-packing_randomized} gives Corollary~\ref{cor:cycle}: a randomized $\mathrm{O}(n^\omega)$-time computation of the minimum length of such a cycle and an $\mathrm{O}(n^{\omega + 1})$-time construction.
The construction bound can be improved accordingly as in Theorem~\ref{thm:S-packing_randomized} with the aid of the binary search when the minimum length is small or large.
Also, combining with Theorem~\ref{thm:S-packing_deterministic} gives a deterministic $\mathrm{O}(nm^3)$-time algorithm.

\section*{Acknowledgments}
We thank the anonymous reviewers of the earlier version for their helpful comments.
In particular, highlighting the usefulness of the finite fields of characteristic $2$ strengthens and simplifies the presentation.

This work was supported by JSPS KAKENHI Grant Numbers 20K19743, 20H00605, and 25H01114, by JST CRONOS Japan Grant Number JPMJCS24K2, and by JST ASPIRE Japan Grant Number JPMJAP2520.

\subsection*{AI Declarations}
The authors used generative AIs to assist in manuscript review and language editing.
All mathematical arguments, references, and suggested revisions were verified by the authors.
The authors are responsible for all descriptions in the submitted manuscript.


\end{document}